# An Algebraic Proof of Thermal Wick's Theorem

BY RUOFAN CHEN

College of Physics and Electronic Engineering, and Center for Computational Sciences, Sichuan Normal University, Chengdu, 610068, China

### Abstract

Every many-body state can be constructed from the vacuum state by iteratively applying the creation operators, therefore the thermal expectation can be transformed into a sum of vacuum expectation where the vacuum Wick's theorem holds. Based on this observation this article gives an algebraic proof of thermal Wick's theorem.

## 1 Introduction

The Wick's theorem is the cornerstone of the Feynmann diagram technique [1, 2, 3, 4, 5, 6, 7, 8], which was proposed by Wick in 1950 with respect to vacuum state. The core statement is that the vacuum expectation value $\langle 0| \cdots |0\rangle$ of any pairs of boson operators $b^\dagger$ and $b$ (here $b^\dagger, b$ are for same state) is equal to the sum of all possible contractions. In each contraction, the operators must be in the original order. For the fermion operators $a^\dagger$ and $a$ (here $a^\dagger, a$ are also for the same state), the rule is the same except that extra sign need to be taken into consideration according to the parity of the number of interchanges needed to bring the operators to the appropriate order.

The proof of the vacuum Wick's theorem relies on the fact the expectation of a normal product, i.e., a product where all the annihilation operators are to the right of the creation operators, is zero. However, such property does not hold for thermal expectation value. Consider the thermal expectation

$$\langle \cdots \rangle = \frac{1}{Z}\text{Tr}[e^{-\beta H} \cdots], \tag{1}$$

where $\beta$ is the inverse temperature, $H = \varepsilon a^\dagger a$ for fermion and $\hat{H} = \varepsilon b^\dagger b$ for boson, and $Z = \text{Tr}\, e^{-\beta H}$ is the partition function. Here we have set $\hbar = k_B = 1$. It is well known that the thermal expectation value of $a^\dagger a$ or $b^\dagger b$, which is a normal product, gives the non-zero Fermi-Dirac or Bose-Einstein distribution.

Due to this reason, althouth the evaluation of the thermal expectation also obeys the Wick's theorem, it can not be proved via the same argument to the vacuum one. The thermal Wick's theorem was first proved by Matsubara [9] for field operator in the thermodynamic limit, whose argument can be found in several textbooks [1, 2]. Latter Gaudin [10, 4] gives a simple proof of the theorem as an operator identity using the commutation rules. In addition, the theorem can be also conveniently proved by path integral formalism [5].

Noticing that the state $|n\rangle$ (take the boson state as an example) can be constructed from the vacuum state as $|n\rangle = \frac{(b^\dagger)^n}{\sqrt{n!}}|0\rangle$, we can transform the thermal expectation value (1) into a sum of vacuum expectation where the Wick's theorem works. Based on such an observation, this article gives an algebraic proof of the thermal Wick's theorem. Sections 2 and 3 are devoted to the proof for fermion, and sections 4 and 5 are for boson.

## 2 "Green's Function" for Fermion

In this section, I shall introduce the basic notations used in the proof for fermion. First, following the standard Green's function terminology, we can define the "time-ordering" operator and the "Green's function". Take a vacuum fermion expectation $\langle 0|aa^\dagger aa^\dagger|0\rangle$ as an example. We can label the operators as $\langle 0|a_1 a_1^\dagger a_2 a_2^\dagger|0\rangle$ and write the expectation as

$$\langle 0|a_1 a_1^\dagger a_2 a_2^\dagger|0\rangle = \chi \langle 0|T a_2 a_1 a_1^\dagger a_2^\dagger|0\rangle, \tag{2}$$

where $T$ is the "time-ordering" operator which means the operators following it should be arranged in the original order. The quantity $\chi$ is the sign according to the parity of number of interchanges needed to bring the operators to the order of the right-handed side. In this example, $a_2$ is brought to the far left by two interchanges, therefore the sign is $\chi = 1$. Employing the vacuum Wick's theorem, the above expectation can be evaluated as

$$\chi\langle 0|Ta_2a_1a_1^\dagger a_2^\dagger|0\rangle = \chi\langle 0|Ta_1a_1^\dagger|0\rangle\langle 0|Ta_2a_2^\dagger|0\rangle - \chi\langle 0|Ta_1a_2^\dagger|0\rangle\langle 0|Ta_2a_1^\dagger|0\rangle$$
$$= \chi[G_{11}G_{22} - G_{12}G_{21}], \tag{3}$$

where $G_{jk} = \langle 0|Ta_ja_k^\dagger|0\rangle$ is the "Green's function". The above expression can be expressed as the determinant of a matrix $G_{jk}$ that

$$\chi\langle 0|Ta_2a_1a_1^\dagger a_2^\dagger|0\rangle = \chi\det\begin{bmatrix} G_{11} & G_{12} \\ G_{21} & G_{22} \end{bmatrix}. \tag{4}$$

It should be emphasized that in the final evaluation of each Green's function, the operators $a$ and $a^\dagger$ need to to be brought back to the original order by the ordering operator $T$. For instance, in the original order $a_1^\dagger$ is on the left of $a_2$, thus

$$G_{21} = \langle 0|Ta_2a_1^\dagger|0\rangle = -\langle 0|a_1^\dagger a_2|0\rangle. \tag{5}$$

Suppose there are now $n$ pairs of $a, a^\dagger$ in the vacuum expectation, using the Green's function terminology defined above, the operators can be always arranged in the "time-ordering" operator as

$$\langle 0|\cdots|0\rangle = \chi\langle 0|Ta_n\cdots a_1a_1^\dagger\cdots a_n^\dagger|0\rangle. \tag{6}$$

Then employing the Wick's theorem yields

$$\langle 0|\cdots|0\rangle = \chi\sum_{\sigma\in S_n}\varepsilon_\sigma G_{1\sigma(1)}\cdots G_{n\sigma(n)}, \tag{7}$$

where $S_n$ is the transformation group of $n$th order and $\varepsilon_\sigma$ is the sign of the permutation $\sigma$. The summation is just the determinant of a matrix $G_{jk}$ that

$$\langle 0|\cdots|0\rangle = \chi\det\begin{bmatrix} G_{11} & \cdots & G_{1n} \\ \vdots & \ddots & \vdots \\ G_{n1} & \cdots & G_{nn} \end{bmatrix}. \tag{8}$$

For conciseness, we denote the matrix $G_{jk}$ as $\hat{G}$, and its $k$th column as $\hat{G}_{(k)}$, then the above determinant can be expressed as

$$\langle 0|\cdots|0\rangle = \chi\det\begin{pmatrix} \hat{G}_{(1)} & \cdots & \hat{G}_{(n)} \end{pmatrix} = \chi\det\hat{G}. \tag{9}$$

Similarly, with the aid of "time-ordering" operator $T$ the thermal expectation can be rewritten as

$$\langle\cdots\rangle = \chi\langle Ta_n\cdots a_1a_1^\dagger\cdots a_n^\dagger\rangle. \tag{10}$$

# 3  Fermion Thermal Wick's theorem

It can be seen from (10) that the thermal expectation of original product $\langle\cdots\rangle$ only differs from the expectation of "time-ordered" product $\langle Ta_n\cdots a_1a_1^\dagger\cdots a_n^\dagger\rangle$ by a sign $\chi$. Therefore if the Wick's theorem holds for the "time-ordered" product, then it is sufficient to prove the thermal Wick's theorem. That is, we can get rid of the sign $\chi$ during the proof.

For fermion, the thermal expectation of a "time-ordered" product can be directly expanded as

$$\langle Ta_n\cdots a_1a_1^\dagger\cdots a_n^\dagger\rangle = \frac{1}{Z}[\langle 0|Ta_n\cdots a_1a_1^\dagger\cdots a_n^\dagger|0\rangle + \alpha\langle 0|a[Ta_n\cdots a_1a_1^\dagger\cdots a_n^\dagger]a^\dagger|0\rangle], \tag{11}$$

where $\alpha = e^{-\beta\varepsilon}$ and we have used the fact that $|1\rangle = \hat{a}^\dagger|0\rangle$. The partition function is simply $Z = 1 + \alpha$. Here we can label the auxiliary $a, a^\dagger$ in the second term of right-handed side as $a_{n+1}, a_{n+1}^\dagger$, then we have

$$\langle Ta_n\cdots a_1a_1^\dagger\cdots a_n^\dagger\rangle = \frac{1}{Z}[\langle 0|Ta_n\cdots a_1a_1^\dagger\cdots a_n^\dagger|0\rangle + \alpha\langle 0|a_{n+1}[Ta_n\cdots a_1a_1^\dagger\cdots a_n^\dagger]a_{n+1}^\dagger|0\rangle]. \tag{12}$$

Note that $a_{n+1}$ and $a_{n+1}^\dagger$ are originally at both ends, thus their position remains unchanged if they are included in the range of "time-ordered" operator $T$. Thus the direct expansion of the thermal expectation can be written as

$$\langle Ta_n \cdots a_1 a_1^\dagger \cdots a_n^\dagger \rangle = \frac{1}{Z}[\langle 0|Ta_n \cdots a_1 a_1^\dagger \cdots a_n^\dagger|0\rangle + \alpha\langle 0|Ta_{n+1}a_n \cdots a_1 a_1^\dagger \cdots a_n^\dagger a_{n+1}^\dagger|0\rangle]. \tag{13}$$

On the other hand, if the thermal Wick's theorem holds, the expectation can be expanded as

$$\langle Ta_n \cdots a_1 a_1^\dagger \cdots a_n^\dagger \rangle = \sum_{\sigma \in S_n} \varepsilon_\sigma \langle Ta_1 a_{\sigma(1)}^\dagger \rangle \cdots \langle Ta_n a_{\sigma(n)}^\dagger \rangle. \tag{14}$$

Our goal is to prove the equivalence of expansions (13) and (14).

Let us first inspect the expectation of a single pair of operators $\langle Ta_1 a_1^\dagger \rangle$, which can be written as

$$\langle Ta_1 a_1^\dagger \rangle = \frac{1}{Z}[\langle 0|Ta_1 a_1^\dagger|0\rangle + \alpha\langle 0|Ta_2 a_1 a_1^\dagger a_2^\dagger|0\rangle] = \frac{1}{Z}\left\{ G_{11} + \alpha \det \begin{bmatrix} G_{11} & G_{12} \\ G_{21} & G_{22} \end{bmatrix} \right\}. \tag{15}$$

It can be seen that $a_2$ is on the left of all $a^\dagger$'s, and $a_2^\dagger$ is on the right of all $a$'s. This means if any index of $G_{jk}$ is 2 then $G_{jk} = 1$, i.e., $G_{12} = G_{21} = G_{22} = 1$. Therefore we have

$$\det \begin{bmatrix} G_{11} & G_{12} \\ G_{21} & G_{22} \end{bmatrix} = \det \begin{bmatrix} G_{11} & 1 \\ 1 & 1 \end{bmatrix} = G_{11} - 1, \tag{16}$$

and

$$\langle Ta_1 a_1^\dagger \rangle = \frac{1}{Z}[G_{11} + \alpha(G_{11} - 1)] = G_{11} - f, \tag{17}$$

where $f = \alpha/(1+\alpha)$. The specific value of $f$ is the Fermi-Dirac distribution $f = (e^{\beta\varepsilon} + 1)^{-1}$. Nevertheless, this specific value is not needed in the proof. If here the $a, a^\dagger$ in $\langle Taa^\dagger \rangle$ are not labeled as $a_1, a_1^\dagger$ but $a_j, a_k^\dagger$, we can still follow the same procedure and obtain

$$\langle Ta_j a_k^\dagger \rangle = G_{jk} - f. \tag{18}$$

Before we proceed to the general situation, let use the expectation of two pairs of operators to illustrate the basic idea. Using direct expansion formula (13) we have

$$\langle Ta_2 a_1 a_1^\dagger a_2^\dagger \rangle = \frac{1}{Z}\left\{ \det \begin{bmatrix} G_{11} & G_{12} \\ G_{21} & G_{22} \end{bmatrix} + \alpha \det \begin{bmatrix} G_{11} & G_{12} & G_{13} \\ G_{21} & G_{22} & G_{23} \\ G_{31} & G_{32} & G_{33} \end{bmatrix} \right\}. \tag{19}$$

Because $a_3$ is on the left of all operators and $a_3^\dagger$ is on the right of all operators, $G_{13}, G_{23}, G_{33}, G_{32}, G_{31}$ are all 1, therefore

$$\langle Ta_2 a_1 a_1^\dagger a_2^\dagger \rangle = \frac{1}{Z}\left\{ \det \begin{bmatrix} G_{11} & G_{12} \\ G_{21} & G_{22} \end{bmatrix} + \alpha \det \begin{bmatrix} G_{11} & G_{12} & 1 \\ G_{21} & G_{22} & 1 \\ 1 & 1 & 1 \end{bmatrix} \right\}. \tag{20}$$

Expanding the second determinant by minors of last row, we have

$$\det \begin{bmatrix} G_{11} & G_{12} & 1 \\ G_{21} & G_{22} & 1 \\ 1 & 1 & 1 \end{bmatrix} = \det\begin{pmatrix} \hat{G}_{(1)} & \hat{G}_{(2)} \end{pmatrix} - \det\begin{pmatrix} \hat{G}_{(1)} & 1 \end{pmatrix} + \det\begin{pmatrix} \hat{G}_{(2)} & 1 \end{pmatrix}. \tag{21}$$

Here the $\begin{pmatrix} \hat{G}_{(1)} & 1 \end{pmatrix}$ means a matrix whose first column is $\hat{G}_{(1)}$ and all the elements of second column is 1, and similar convention applies for $\begin{pmatrix} \hat{G}_{(2)} & 1 \end{pmatrix}$. We can move the column $\hat{G}_{(2)}$ in $\det\begin{pmatrix} \hat{G}_{(2)} & 1 \end{pmatrix}$ to the second column, which yields a minus sign due to the skew symmetry of the determinant. That is, $\det\begin{pmatrix} \hat{G}_{(2)} & 1 \end{pmatrix} = -\det\begin{pmatrix} 1 & \hat{G}_{(2)} \end{pmatrix}$, then we have

$$\langle Ta_2 a_1 a_1^\dagger a_2^\dagger \rangle = \det\begin{pmatrix} \hat{G}_{(1)} & \hat{G}_{(2)} \end{pmatrix} - f\det\begin{pmatrix} 1 & \hat{G}_{(2)} \end{pmatrix} - f\det\begin{pmatrix} \hat{G}_{(1)} & 1 \end{pmatrix}. \tag{22}$$

Denote $\hat{B}_k$ as a matrix obtained by replacing the elements of $k$th column of $\hat{G}$ by 1, e.g., here $\hat{B}_1 = \begin{pmatrix} 1 & \hat{G}_{(1)} \end{pmatrix}$ and $\hat{B}_2 = \begin{pmatrix} \hat{G}_{(1)} & 1 \end{pmatrix}$. With this convention, we can write

$$\langle Ta_2 a_1 a_1^\dagger a_2^\dagger \rangle = \det\hat{G} - f(\det\hat{B}_1 + \det\hat{B}_2). \tag{23}$$

On the other hand, employing the thermal Wick's theorem we shall have

$$
\begin{aligned}
\langle Ta_2 a_1 a_1^\dagger a_2^\dagger \rangle &= \langle Ta_1 a_1^\dagger \rangle \langle Ta_2 a_2^\dagger \rangle - \langle Ta_1 a_2^\dagger \rangle \langle Ta_2 a_1^\dagger \rangle \\
&= \det \begin{bmatrix} G_{11}-f & G_{12}-f \\ G_{21}-f & G_{22}-f \end{bmatrix} \\
&= \det\left( \hat{G}_{(1)}-f \ \ \hat{G}_{(2)}-f \right),
\end{aligned}
\tag{24}
$$

where $\hat{G}_{(k)}-f$ means a column obtained by substrating $f$ from every element of $\hat{G}_{(k)}$. Since the determinant is multi-linear of columns, expanding the first column of above determinant gives

$$
\det\left( \hat{G}_{(1)}-f \ \ \hat{G}_{(2)}-f \right) = \det\left( \hat{G}_{(1)} \ \ \hat{G}_{(2)}-f \right) - \det\left( f \ \ \hat{G}_{(2)}-f \right).
\tag{25}
$$

Since the determinant is invariant under elementary transformation of type II, we can add the first column of $\det\begin{bmatrix} f & \hat{G}_{(2)}-f \end{bmatrix}$ to the second column, which yields

$$
\det\left( f \ \ \hat{G}_{(2)}-f \right) = \det\left( f \ \ \hat{G}_{(2)} \right).
\tag{26}
$$

Now expand the second column of the first determinant in the right-handed side of (25), we shall finally have

$$
\begin{aligned}
\det\left( \hat{G}_{(1)}-f \ \ \hat{G}_{(2)}-f \right) &= \det\left( \hat{G}_{(1)} \ \ \hat{G}_{(2)} \right) - \det\left( \hat{G}_{(1)} \ \ f \right) - \det\left( f \ \ \hat{G}_{(2)} \right) \\
&= \det\hat{G} - f(\det\hat{B}_1 + \det\hat{B}_2),
\end{aligned}
\tag{27}
$$

which coincides with (23). Therefore the thermal Wick's theorem holds for product of two pairs of operators.

Now let us proceed to the general situation where there are $n > 2$ pairs of operators. The direct expansion formula (13) gives

$$
\langle Ta_n \cdots a_1 a_1^\dagger \cdots a_n^\dagger \rangle = \frac{1}{Z}\left\{ \det \begin{bmatrix} G_{11} & \cdots & G_{1n} \\ \vdots & \ddots & \vdots \\ G_{n1} & \cdots & G_{nn} \end{bmatrix} + \alpha \det \begin{bmatrix} G_{11} & \cdots & G_{1n} & 1 \\ \vdots & & \vdots & \vdots \\ G_{n1} & \cdots & G_{nn} & 1 \\ 1 & \cdots & 1 & 1 \end{bmatrix} \right\}.
\tag{28}
$$

Expand the second determinant by minors of last row and swap the column consisting of all 1 to the proper position, we shall have

$$
\langle Ta_n \cdots a_1 a_1^\dagger \cdots a_n^\dagger \rangle = \det\hat{G} - f\sum_{k=1}^{n} \det\hat{B}_k.
\tag{29}
$$

At the same time, employing the thermal Wick's theorem we have

$$
\langle Ta_n \cdots a_1 a_1^\dagger \cdots a_n^\dagger \rangle = \det\left( \hat{G}_{(1)}-f \ \ \cdots \ \ \hat{G}_{(n)}-f \right).
\tag{30}
$$

Expanding the first column yields

$$
\det\left( \hat{G}_{(1)}-f \ \ \cdots \ \ \hat{G}_{(n)}-f \right) = \det\left( \hat{G}_{(1)} \ \ \hat{G}_{(2)}-f \ \ \cdots \ \ \hat{G}_{(n)}-f \right) - \det\left( f \ \ \hat{G}_{(2)}-f \ \ \cdots \ \ \hat{G}_{(n)}-f \right).
\tag{31}
$$

Add the first column to the rest columns in the second determinant of right-handed side above, we have

$$
\det\left( \hat{G}_{(1)}-f \ \ \cdots \ \ \hat{G}_{(n)}-f \right) = \det\left( \hat{G}_{(1)} \ \ \hat{G}_{(2)}-f \ \ \cdots \ \ \hat{G}_{(n)}-f \right) - \det\left( f \ \ \hat{G}_{(2)} \ \ \cdots \ \ \hat{G}_{(n)} \right).
\tag{32}
$$

Repeat this procedure, we shall finally obtain

$$
\begin{aligned}
\langle Ta_n \cdots a_1 a_1^\dagger \cdots a_n^\dagger \rangle &= \det\left( \hat{G}_{(1)} \ \ \cdots \ \ \hat{G}_{(n)} \right) - \det\left( f \ \ \hat{G}_{(2)} \ \ \cdots \ \ \hat{G}_{(n)} \right) - \cdots \\
&\quad - \det\left( \hat{G}_{(1)} \ \ \cdots \ \ f \ \ \cdots \ \ \hat{G}_{(n)} \right) - \cdots - \det\left( \hat{G}_{(1)} \ \ \cdots \ \ \hat{G}_{(k)} \ \ \cdots \ \ f \right).
\end{aligned}
\tag{33}
$$

which coincides with (29), and this complete the proof of fermion thermal Wick's theorem.

## 4 Full Permutation of A Square Matrix

For boson, the "time-ordering" operator and the "Green's function" can be defined in a same way to those of fermion, except that there is no sign issue. Suppose there are $n$ pairs of operators in the vacuum expectation, using the "time-ordering" operator we can write it as

$$
\langle 0 | \cdots | 0 \rangle = \langle 0 | Tb_n \cdots b_1 b_1^\dagger \cdots b_n^\dagger | 0 \rangle.
\tag{34}
$$

Expanding it via vacuum Wick's theorem, we have

$$\langle 0|Tb_n\cdots b_1 b_1^\dagger \cdots b_n^\dagger|0\rangle \;=\; \sum_{\sigma\in S_n} G_{1\sigma(1)}\cdots G_{n\sigma(n)}, \tag{35}$$

where the "time-ordered Green's function" is defined as

$$G_{jk} = \langle 0|Tb_j b_k^\dagger|0\rangle. \tag{36}$$

Now define the full permutation of a square matrix as

$$\mathrm{fp}\,\hat{G} = \mathrm{fp}\begin{bmatrix} G_{11} & \cdots & G_{1n} \\ \vdots & \ddots & \vdots \\ G_{n1} & \cdots & G_{nn} \end{bmatrix} = \sum_{\sigma\in S_n} G_{1\sigma(1)}\cdots G_{n\sigma(n)}, \tag{37}$$

then (35) can be expressed as

$$\langle 0|Tb_n\cdots b_1 b_1^\dagger \cdots b_n^\dagger|0\rangle = \mathrm{fp}\,\hat{G}. \tag{38}$$

The definition of the full permutation (37) is somehow similar to the definition of the determinant, except that there is no sign factor $\varepsilon_\sigma$. According to this definition, the full permutation is multi-linear of rows or columns. It is also symmetric with respect to rows or columns, i.e., it is invariant under swapping of any two rows or columns.

We will need an expansion formula for the full permutation, which is analogous to the general Laplace expansion theorem for the determinant. For a given $n\times n$ matrix $\hat{G}$, denote its $k\times k$ submatrix formed from rows $i_1,\dots,i_k$ and columns $j_1,\dots,j_k$ as (this corresponding to the minor of order $k$ in determinant terminology)

$$M\begin{pmatrix} i_1 & \cdots & i_k \\ j_1 & \cdots & j_k \end{pmatrix}, \tag{39}$$

and the $(n-k)\times(n-k)$ submatrix obtained by removing rows $i_1,\dots,i_k$ and columns $j_1,\dots,j_k$ as (this corresponding to the cofactor in determinant terminology)

$$\overline{M}\begin{pmatrix} i_1 & \cdots & i_k \\ j_1 & \cdots & j_k \end{pmatrix}. \tag{40}$$

Then the full permutation of $G$ can be expanded as

$$\mathrm{fp}\,\hat{G} = \sum_{1\leqslant j_1<\cdots<j_k\leqslant n} \mathrm{fp}\,M\begin{pmatrix} i_1 & \cdots & i_k \\ j_1 & \cdots & j_k \end{pmatrix} \cdot \mathrm{fp}\,\overline{M}\begin{pmatrix} i_1 & \cdots & i_k \\ j_1 & \cdots & j_k \end{pmatrix}. \tag{41}$$

This expansion formula can be derived as follows. First, since the full permutation is symmetric of rows and columns, without losing generality we can set $i_1=1,\dots,i_k=k$, i.e., consider the expansion via first $k$ rows. By definition, the full permutation is

$$\mathrm{fp}\,\hat{G} = \sum_{\sigma\in S_n} G_{1\sigma(1)}\cdots G_{k\sigma(k)} \times G_{k+1,\sigma(k+1)}\cdots G_{n\sigma(n)}. \tag{42}$$

For a given set of $j_1,\dots,j_k$, the elements chosen in first $k$ rows gives the product $G_{1j_1}\cdots G_{kj_k}$, we can permute the column indices $j_1,\dots,j_k$ in all the possibilities and sum up the corresponding product, which would yield the full permutation $\mathrm{fp}\,M\begin{pmatrix} 1 & \cdots & k \\ j_1 & \cdots & j_k \end{pmatrix}$. At the same time, the summation of all permutation in last $n-k$ rows sums up as $\mathrm{fp}\,\overline{M}\begin{pmatrix} 1 & \cdots & k \\ j_1 & \cdots & j_k \end{pmatrix}$, together we obtain $\mathrm{fp}\,M\begin{pmatrix} 1 & \cdots & k \\ j_1 & \cdots & j_k \end{pmatrix} \cdot \mathrm{fp}\,\overline{M}\begin{pmatrix} 1 & \cdots & k \\ j_1 & \cdots & j_k \end{pmatrix}$. Then we can choose another set of $j_1,\dots,j_k$ and repeat this procedure, after all the combiantions of column indices $j_1,\dots,j_k$ are traversed, the desire expansion formula (41) is obtained.

## 5  Boson Thermal Wick's Theorem

Suppose there are $n$ pairs of operators, the boson thermal expectation can be expanded directly as

$$\langle Tb_n\cdots b_1 b_1^\dagger \cdots b_n^\dagger\rangle = \frac{1}{Z}\sum_{m=0}^{\infty} \frac{\alpha^m}{m!} \langle 0|b^m[Tb_n\cdots b_1 b_1^\dagger \cdots b_n^\dagger](b^\dagger)^m|0\rangle, \tag{43}$$

where $\alpha = e^{-\beta\varepsilon}$ and we have used the fact $|m\rangle = \frac{(b^\dagger)^m}{\sqrt{m!}}|0\rangle$. The partition function is $Z = \sum_{m=0}^{\infty}\alpha^m = (1-\alpha)^{-1}$. As in the fermion case, the auxiliary $b$'s and $b^\dagger$'s are at both ends and their position remains the same when included in the range of "time-ordering" operator, thus the above expression can be labled as

$$\langle Tb_n\cdots b_1 b_1^\dagger\cdots b_n^\dagger\rangle = \frac{1}{Z}\sum_{m=0}^{\infty}\frac{\alpha^m}{m!}\langle 0|Tb_{n+m}\cdots b_{n+1}b_n\cdots b_1 b_1^\dagger\cdots b_n^\dagger b_{n+1}^\dagger\cdots b_{n+m}^\dagger|0\rangle. \tag{44}$$

Let us first inspect the thermal expectation of a single pair of operators that

$$\langle Tb_1 b_1^\dagger\rangle = \frac{1}{Z}\sum_{m=0}^{\infty}\frac{\alpha^m}{m!}\langle 0|Tb_{m+1}\cdots b_1 b_1^\dagger\cdots b_m^\dagger|0\rangle. \tag{45}$$

Because $b_2,\ldots,b_{m+1}$ are on the left of all $b^\dagger$'s and $b_2^\dagger,\ldots,b_m^\dagger$ are on the right of all $b$'s, we have $G_{ij}=1$ if $i\geqslant 2$ or $j\geqslant 2$. Employing the vacuum Wick's theorem on $m$th order term, we can write

$$\frac{\alpha^m}{m!}\langle 0|Tb_{m+1}\cdots b_1 b_1^\dagger\cdots b_m^\dagger|0\rangle = \frac{\alpha^m}{m!}\mathrm{fp}\,\hat{\mathscr{G}}^{(m)}, \tag{46}$$

where $\hat{\mathscr{G}}^{(m)}$ is a $(m+1)\times(m+1)$ matrix full of 1 except an element at the upper left corner that

$$\hat{\mathscr{G}}^{(m)} = \begin{bmatrix} G_{11} & 1 & \cdots & 1 \\ 1 & \ddots & & 1 \\ \vdots & & \ddots & \vdots \\ 1 & 1 & \cdots & 1 \end{bmatrix}. \tag{47}$$

Using the expansion formula (41), the full permutation of $\hat{\mathscr{G}}^{(m)}$ can be expanded via first row that

$$\mathrm{fp}\,\hat{\mathscr{G}}^{(m)} = \sum_{j=1}^{m+1}\mathrm{fp}\,M\begin{pmatrix}1\\j\end{pmatrix}\cdot\mathrm{fp}\,\overline{M}\begin{pmatrix}1\\j\end{pmatrix},$$

where $M\begin{pmatrix}1\\j\end{pmatrix}$ is $[\,G_{11}\,]$ if $j=1$, or $[\,1\,]$ otherwise, and thus $\mathrm{fp}\,M\begin{pmatrix}1\\j\end{pmatrix}=G_{11}$ if $j=1$ and $\mathrm{fp}\,M\begin{pmatrix}1\\j\end{pmatrix}=1$ otherwise. The submatrix $\overline{M}\begin{pmatrix}1\\j\end{pmatrix}$ is a $m\times m$ matrix full of 1, therefore its full permutation is simply $\mathrm{fp}\,\overline{M}\begin{pmatrix}1\\j\end{pmatrix}=m!$. Finally, we find that (45) is

$$\begin{aligned}\langle Tb_1 b_1^\dagger\rangle &= \frac{1}{Z}\sum_{m=0}^{\infty}\frac{\alpha^m}{m!}(G_{11}+m)\cdot m! \\ &= G_{11}+f,\end{aligned} \tag{48}$$

where $f = \frac{1}{Z}\sum_{m=0}^{\infty}\alpha^m m = \alpha/(1-\alpha)$. The specific value of $f$ is the Bose-Einstein distribution $f = (e^{\beta\varepsilon}-1)^{-1}$, and as in the fermion case this specific value is not needed in the proof.

Again, let us use the expectation of two pairs of operators to illustrate the basic idea. The thermal expectation can be directly expanded as

$$\langle Tb_2 b_1 b_1^\dagger b_2^\dagger\rangle = \frac{1}{Z}\sum_{m=0}^{\infty}\frac{\alpha^m}{m!}\langle 0|b_{m+2}\cdots b_2 b_1 b_1^\dagger b_2^\dagger\cdots b_{m+2}^\dagger|0\rangle. \tag{49}$$

Employing the vacuum Wick's theorem on the $m$th order term gives

$$\frac{\alpha^m}{m!}\langle 0|Tb_{m+1}\cdots b_2 b_1 b_1^\dagger b_2^\dagger\cdots b_m^\dagger|0\rangle = \frac{\alpha^m}{m!}\,\mathrm{fp}\,\hat{\mathscr{G}}^{(m)}, \tag{50}$$

where $\hat{\mathscr{G}}^{(m)}$ is a $(m+2)\times(m+2)$ matrix full of 1 except a $2\times 2$ submatrix at the upper left corner that

$$\hat{\mathscr{G}}^{(m)} = \begin{bmatrix} G_{11} & G_{12} & 1 & \cdots & 1 \\ G_{21} & G_{22} & 1 & \cdots & 1 \\ 1 & \cdots & 1 & \cdots & 1 \\ \vdots & \ddots & \vdots & \ddots & \vdots \\ 1 & \cdots & 1 & \cdots & 1 \end{bmatrix}, \tag{51}$$

here we shall denote the submatrix $\begin{bmatrix} G_{11} & G_{12} \\ G_{21} & G_{22} \end{bmatrix}$ as $\hat{G} = \begin{pmatrix} \hat{G}_{(1)} & \hat{G}_{(2)} \end{pmatrix}$. Expanding the full permutation fp $\hat{\mathscr{G}}^{(m)}$ by the first two rows via (41) gives

$$\text{fp}\,\hat{\mathscr{G}}^{(m)} = \sum_{1 \leqslant j_1 < j_2 \leqslant m+2} \text{fp}\, M \begin{pmatrix} 1 & 2 \\ j_1 & j_2 \end{pmatrix} \cdot \text{fp}\, \overline{M} \begin{pmatrix} 1 & 2 \\ j_1 & j_2 \end{pmatrix}. \tag{52}$$

It is easy to see $\overline{M} \begin{pmatrix} 1 & 2 \\ j_1 & j_2 \end{pmatrix}$ is a $m \times m$ matrix full of 1, whose full permutation is simply fp $\overline{M} \begin{pmatrix} 1 & 2 \\ j_1 & j_2 \end{pmatrix} = m!$. The $2 \times 2$ submatrix $M \begin{pmatrix} 1 & 2 \\ j_1 & j_2 \end{pmatrix}$ formed from the first two rows and columns $j_1, j_2$, and traversing all the possible $j_1$, $j_2$ sums up as

$$\frac{1}{m!} \text{fp}\,\hat{\mathscr{G}}^{(m)} = \text{fp}\begin{pmatrix} \hat{G}_{(1)} & \hat{G}_{(2)} \end{pmatrix} + C_m^1 \text{fp}\begin{pmatrix} 1 & \hat{G}_{(2)} \end{pmatrix} + C_m^1 \text{fp}\begin{pmatrix} \hat{G}_{(1)} & 1 \end{pmatrix} + C_m^2 \text{fp}\begin{pmatrix} 1 & 1 \end{pmatrix}, \tag{53}$$

where $C_m^k$ is the usual combiantorial number for $m \geqslant k$ and it is set to zero if $m < k$. Here we have used the symmetry fp$\begin{pmatrix} \hat{G}_{(1)} & 1 \end{pmatrix} = \text{fp}\begin{pmatrix} 1 & \hat{G}_{(2)} \end{pmatrix}$, and $\begin{pmatrix} 1 & 1 \end{pmatrix}$ is a $2 \times 2$ matrix whose first and second columns are all 1. Summing up from $m = 0$ to $\infty$, the first three terms yields

$$\frac{1}{Z}\sum_{m=0}^{\infty} \alpha^m \cdot \text{fp}\begin{pmatrix} \hat{G}_{(1)} & \hat{G}_{(2)} \end{pmatrix} = \text{fp}\begin{pmatrix} \hat{G}_{(1)} & \hat{G}_{(2)} \end{pmatrix}, \tag{54}$$

$$\frac{1}{Z}\sum_{m=0}^{\infty} C_m^1 \alpha^m \cdot \text{fp}\begin{pmatrix} 1 & \hat{G}_{(2)} \end{pmatrix} = f\,\text{fp}\begin{pmatrix} 1 & \hat{G}_{(2)} \end{pmatrix}, \tag{55}$$

and

$$\frac{1}{Z}\sum_{m=0}^{\infty} C_m^1 \alpha^m \cdot \text{fp}\begin{pmatrix} \hat{G}_{(1)} & 1 \end{pmatrix} = f\,\text{fp}\begin{pmatrix} \hat{G}_{(1)} & 1 \end{pmatrix}. \tag{56}$$

The last term can be conveniently evaluated in a calculus manner, and a more algebraic method would be given in the appendix. Recall that $Z = \sum_{m=0}^{\infty} \alpha^m = (1-\alpha)^{-1}$, we have

$$\frac{1}{Z}\sum_{m=0}^{\infty} C_m^2 \alpha^m = \frac{\alpha^2}{Z}\frac{\mathrm{d}^2}{\mathrm{d}\alpha^2}\left[\frac{1}{2!}\sum_{m=0}^{\infty} \alpha^m\right] = \frac{\alpha^2}{2!Z}\frac{\mathrm{d}^2 Z}{\mathrm{d}\alpha^2} = \frac{\alpha^2}{(1-\alpha)^2} = f^2, \tag{57}$$

and thus the last term yields

$$\frac{1}{Z}\sum_{m=0}^{\infty} C_m^2 \alpha^m \cdot \text{fp}\begin{pmatrix} 1 & 1 \end{pmatrix} = f^2 \text{fp}\begin{pmatrix} 1 & 1 \end{pmatrix}. \tag{58}$$

Finally, the direct expansion along with the vacuum Wick's theorem gives

$$\langle Tb_2 b_1 b_1^\dagger b_2^\dagger \rangle = \text{fp}\begin{pmatrix} \hat{G}_{(1)} & \hat{G}_{(2)} \end{pmatrix} + f\,\text{fp}\begin{pmatrix} 1 & \hat{G}_{(2)} \end{pmatrix} + f\,\text{fp}\begin{pmatrix} \hat{G}_{(1)} & 1 \end{pmatrix} + f^2 \text{fp}\begin{pmatrix} 1 & 1 \end{pmatrix}. \tag{59}$$

If the thermal Wick's theorem holds, we have the expansion

$$\langle Tb_2 b_1 b_1^\dagger b_2^\dagger \rangle = \langle Tb_1 b_1^\dagger \rangle \langle Tb_2 b_2^\dagger \rangle + \langle Tb_1 b_2^\dagger \rangle \langle Tb_2 b_1^\dagger \rangle$$
$$= \text{fp}\begin{bmatrix} G_{11}+f & G_{12}+f \\ G_{21}+f & G_{22}+f \end{bmatrix}. \tag{60}$$

Using the multi-linear property of the full permutation, the above expression can be expanded as

$$\text{fp}\begin{bmatrix} G_{11}+f & G_{12}+f \\ G_{21}+f & G_{22}+f \end{bmatrix} = \text{fp}\begin{bmatrix} G_{11} & G_{12} \\ G_{21} & G_{22} \end{bmatrix} + f\begin{bmatrix} 1 & G_{12} \\ 1 & G_{22} \end{bmatrix} + f\begin{bmatrix} G_{11} & 1 \\ G_{21} & 1 \end{bmatrix} + f^2\begin{bmatrix} 1 & 1 \\ 1 & 1 \end{bmatrix},$$

which coincides with (59). Therefore the thermal Wick's theorem holds for two pairs of operators.

Now proceed to the general case where there are $n$ pairs of operators. Direct expansion gives

$$\langle Tb_n \cdots b_1 b_1^\dagger \cdots b_n^\dagger \rangle = \frac{1}{Z}\sum_{m=0}^{\infty} \frac{\alpha^m}{m!} \langle 0| b_{n+m} \cdots b_1 b_1^\dagger \cdots b_{n+m}^\dagger |0\rangle$$
$$= \frac{1}{Z}\sum_{m=0}^{\infty} \frac{\alpha^m}{m!} \text{fp}\,\hat{\mathscr{G}}^{(m)}, \tag{61}$$

where $\hat{\mathscr{G}}^{(m)}$ is $(n+m)\times(n+m)$ matrix that

$$\hat{\mathscr{G}}^{(m)}=\begin{bmatrix} G_{11} & \cdots & G_{1n} & 1 & \cdots & 1 \\ \vdots & \ddots & \vdots & \vdots & \ddots & \vdots \\ G_{n1} & \cdots & G_{nn} & 1 & \cdots & 1 \\ 1 & \cdots & 1 & 1 & \cdots & 1 \\ \vdots & 1 & \vdots & \vdots & \ddots & 1 \\ 1 & \cdots & 1 & 1 & 1 & 1 \end{bmatrix}. \tag{62}$$

Basically, the upper left block of $\hat{\mathscr{G}}^{(m)}$ is the matrix $\hat{G}$ and the rest elements are all 1. Employing (41), $\hat{\mathscr{G}}^{(m)}$ can be expanded by first $n$ rows as

$$\hat{\mathscr{G}}^{(m)}=\sum_{1\leqslant j_1<\cdots<j_n\leqslant n+m}\mathrm{fp}\,M\begin{pmatrix} 1 & \cdots & n \\ j_1 & \cdots & j_n \end{pmatrix}\cdot\mathrm{fp}\,\overline{M}\begin{pmatrix} 1 & \cdots & n \\ j_1 & \cdots & j_n \end{pmatrix}, \tag{63}$$

where $\mathrm{fp}\,\overline{M}\begin{pmatrix} 1 & \cdots & n \\ j_1 & \cdots & j_n \end{pmatrix}$ simply gives $m!$ which cancels the factor $\frac{1}{m!}$ in (61).

Now inspect the full permutation of $M\begin{pmatrix} 1 & \cdots & n \\ j_1 & \cdots & j_n \end{pmatrix}$. If all $j_1,\ldots,j_n\leqslant n$, then there is only one possibility that $j_1=1,j_2=2,\ldots,j_n=n$, we have $M\begin{pmatrix} 1 & \cdots & n \\ j_1 & \cdots & j_n \end{pmatrix}=\hat{G}$. Denote $\hat{B}_{n_1}$ as the matrix obtained from $\hat{G}$ by replacing $n_1$th column by 1. If one integer, say $1\leqslant n_1\leqslant n$, is not in $j_1,\ldots,j_n$, then there are totally $C_m^1$ possibilities to choose $n<j_n\leqslant n+m$. The $n$th column of $\mathrm{fp}\,M\begin{pmatrix} 1 & \cdots & n \\ j_1 & \cdots & j_n \end{pmatrix}$ would all 1, and we can interchange this column to $n_1$th column, summing up all these terms we have $C_m^1\mathrm{fp}\,\hat{B}_{n_1}$. Then traversing $1\leqslant n_1\leqslant n$ gives $\sum_{1\leqslant n_1\leqslant n}C_m^1\mathrm{fp}\,\hat{B}_{n_1}$. Similarly, if two integers, say $1\leqslant n_1<n_2\leqslant n$, are not in $j_1,\ldots,j_n$, then it gives $\sum_{1\leqslant n_1<n_2\leqslant n}C_m^2\hat{B}_{n_1n_2}$, where $\hat{B}_{n_1n_2}$ is obtained from $\hat{G}$ by replacing its $n_1$th and $n_2$th columns to 1. This procedure can go on, if $k$ columns are replaced by 1, we have $\sum_{1\leqslant n_1<\cdots<n_k\leqslant n}C_m^k\hat{B}_{n_1\cdots n_k}$. Since

$$\frac{1}{Z}\sum_{m=0}C_m^k\alpha^m=\frac{\alpha^k}{k!Z}\frac{\mathrm{d}^kZ}{\mathrm{d}\alpha^k}=f^k, \tag{64}$$

we can conclude that after summing up from $m=0$ to $\infty$, the coefficient of $\mathrm{fp}\,\hat{B}_{n_1}$ is $f$, and that of $\mathrm{fp}\,\hat{B}_{n_1n_2}$ is $f^2$, and so on. In summary, in (63) every $M\begin{pmatrix} 1 & \cdots & n \\ j_1 & \cdots & j_n \end{pmatrix}$ is obtained from $\hat{G}$ by replacing some columns by 1, and if $k$ columns are replaced then after summation of $m$ it is assigned a coefficient $f^k$. That is, (61) can be written as

$$\langle Tb_n\cdots b_1 b_1^{\dagger}\cdots b_n^{\dagger}\rangle=\mathrm{fp}\,\hat{G}+f\sum_{1\leqslant k\leqslant n}\mathrm{fp}\,\hat{B}_k+f^2\sum_{1\leqslant k_1<k_2\leqslant n}\mathrm{fp}\,\hat{B}_{k_1k_2}+\cdots+f^n\sum_{1\leqslant k_1<\cdots<k_n\leqslant n}\mathrm{fp}\,\hat{B}_{k_1\cdots k_n}, \tag{65}$$

note that in the last term all columns are replaced by 1, i.e., $\hat{B}_{k_1\cdots k_n}$ is a $n\times n$ matrix full of 1.

The thermal Wick's theorem gives the expansion

$$\begin{aligned} \langle Tb_n\cdots b_1 b_1^{\dagger}\cdots b_n^{\dagger}\rangle &= \mathrm{fp}\begin{bmatrix} G_{11}+f & \cdots & G_{1n}+f \\ \vdots & \ddots & \vdots \\ G_{n1}+f & \cdots & G_{nn}+f \end{bmatrix} \\ &= \mathrm{fp}\begin{pmatrix} \hat{G}_{(1)}+f & \cdots & \hat{G}_{(n)}+f \end{pmatrix}. \end{aligned} \tag{66}$$

Using the multi-linearity to expand the above full permutation by its columns, we have

$$\mathrm{fp}\begin{pmatrix} \hat{G}_{(1)}+f & \cdots & \hat{G}_{(n)}+f \end{pmatrix}=\mathrm{fp}\begin{pmatrix} \hat{G}_{(1)} & \hat{G}_{(2)}+f & \cdots & \hat{G}_{(n)}+f \end{pmatrix}+f\mathrm{fp}\begin{pmatrix} 1 & \hat{G}_{(2)}+f & \cdots & \hat{G}_{(n)}+f \end{pmatrix}. \tag{67}$$

This expansion process can go on. For example, in next step we can expand

$$\mathrm{fp}\begin{pmatrix} \hat{G}_{(1)} & \hat{G}_{(2)}+f & \cdots & \hat{G}_{(n)}+f \end{pmatrix}=\mathrm{fp}\begin{pmatrix} \hat{G}_{(1)} & \hat{G}_{(2)} & \cdots & \hat{G}_{(n)}+f \end{pmatrix}+f\mathrm{fp}\begin{pmatrix} \hat{G}_{(1)} & 1 & \cdots & \hat{G}_{(n)}+f \end{pmatrix}, \tag{68}$$

and

$$f\mathrm{fp}\begin{pmatrix} 1 & \hat{G}_{(2)}+f & \cdots & \hat{G}_{(n)}+f \end{pmatrix}=f\mathrm{fp}\begin{pmatrix} 1 & \hat{G}_{(2)} & \cdots & \hat{G}_{(n)}+f \end{pmatrix}+f^2\mathrm{fp}\begin{pmatrix} 1 & 1 & \cdots & \hat{G}_{(n)} \end{pmatrix}. \tag{69}$$

The whole process is sort of like the binominal expansion, in the end we shall find that if $k$ columns in a matrix are 1, then the coefficient of its full permutation is $f^k$. Summing up all the possibilities we have the same expression to (65), which completes the proof.

*Acknowledgment*. This work is supported by National Natural Science Foundation of China under Grant No. 12104328.

# Appendix

In section 5, the term $\frac{1}{Z}\sum_{m=0} C_m^k \alpha^m$ is evaluted in a calculus manner, and a more algebraic way is discussed here. The partition function can be written as a geometric series that $Z = \sum_{m=0}^{\infty} \alpha^m$, and thus

$$Z - \alpha Z = \sum_{m=0}^{\infty} \alpha^m - \sum_{m=1}^{\infty} \alpha^m = 1, \tag{70}$$

which gives $Z = (1 - \alpha)^{-1}$. Define a series $F_1 = \sum_{m=0}^{\infty} C_m^1 \alpha^m = \sum_{m=0}^{\infty} m \alpha^m$, which is related to $f$ by $f = \frac{F_1}{Z}$. It can be seen that

$$F_1 - \alpha F_1 = \sum_{m=1}^{\infty} \alpha^m = \alpha Z, \tag{71}$$

which gives

$$f = \frac{F_1}{Z} = \frac{\alpha}{(1 - \alpha)}. \tag{72}$$

Define $F_2 = \sum_{m=0}^{\infty} C_m^2 \alpha^m$, then we have

$$F_2 - \alpha F_2 = \alpha \sum_{m=0}^{\infty} m \alpha^m = \alpha F_1, \tag{73}$$

that is, $F_2 = \frac{\alpha}{(1 - \alpha)} F_1 = f F_1$. Now define $F_k = \sum_{m=0}^{\infty} C_m^k \alpha^m$ for $k \geqslant 2$, then we have

$$F_{k+1} - \alpha F_{k+1} = \alpha \sum_{m=0}^{\infty} C_m^k \alpha^m = \alpha F_k, \tag{74}$$

which gives a recursive relation

$$F_{k+1} = \frac{\alpha}{1 - \alpha} F_k = f F_k.$$

Finally we have the desire result

$$\frac{1}{Z} \sum_{m=0}^{\infty} C_m^k \alpha^k = \frac{F_k}{Z} = f^{k-1} \frac{F_1}{Z} = f^k. \tag{75}$$